\documentclass{pasa}%

\title[Grand challenges in protoplanetary disc modelling]{Grand challenges in protoplanetary disc modelling}
\author[T. J. Haworth et al.]{Thomas J. Haworth$^{1,2}$\thanks{thaworth@ast.cam.ac.uk}, John D. Ilee$^1$\thanks{jdilee@ast.cam.ac.uk}, Duncan H. Forgan$^3$\thanks{dhf3@st-andrews.ac.uk}, Stefano Facchini$^{4}$, Daniel J. Price$^5$ and \smallskip \\ 
\textit{Community authors}\thanks{This paper was coordinated and written by the first 5 authors: Haworth, Ilee, Forgan, Facchini and Price. The additional ``community authors'', presented alphabetically, made valuable contributions that helped to inform the manuscript.}: Dominika M. Boneberg$^1$, Richard A. Booth$^1$, Cathie J. Clarke$^1$, Jean-Fran{\c c}ois Gonzalez$^6$, \\ Mark A. Hutchison$^7$, Inga Kamp$^8$, Guillaume Laibe$^3$, Wladimir Lyra$^{9,10}$, Farzana Meru$^1$, Subhanjoy Mohanty$^2$, \\ Olja Pani\'{c}$^{1, 11}$\thanks{Royal Society Dorothy Hodgkin Fellow},  Ken Rice$^{12}$, Takeru Suzuki$^{13, 14}$, Richard Teague$^{15}$, Catherine Walsh$^{11, 16}$, Peter Woitke$^2$ \\
\affil{$^1$ Institute of Astronomy, Madingley Road, Cambridge, CB3 0HA, UK}%
\affil{$^2$ Astrophysics Group, Imperial College London, Blackett Laboratory, Prince Consort Road, London SW7 2AZ, UK}
\affil{$^3$ School of Physics and Astronomy, University of St Andrews, North Haugh, St Andrews KY16 9SS, UK}%
\affil{$^4$ Max-Planck-Institut f\"ur Extraterrestrische Physik, Giessenbachstrasse 1, 85748 Garching, Germany}%
\affil{$^5$ Monash Centre for Astrophysics and School of Physics and Astronomy, Monash University, Clayton, VIC 3800, Australia}
\affil{$^6$ Univ Lyon, Univ Lyon1, Ens de Lyon, CNRS, Centre de Recherche Astrophysique de Lyon UMR5574, F-69230, Saint-Genis-Laval, France}
\affil{$^7$ Centre for Astrophysics and Supercomputing, Swinburne University of Technology, Hawthorn, VIC 3122, Australia}
\affil{$^{8}$ Kapteyn Astronomical Institute, Postbus 800, 9700 AV Groningen, The Netherlands}
\affil{$^9$ Department of Physics and Astronomy, California State University
Northridge, 18111 Nordhoff St, Northridge, CA 91330, USA}
\affil{$^{10}$ Jet Propulsion Laboratory, California Institute of Technology, 4800 Oak Grove Drive, Pasadena, CA, 91109, USA}
\affil{$^{11}$ School of Physics and Astronomy, University of Leeds, Leeds, LS2 9JT, UK}
\affil{$^{12}$ Institute for Astronomy, University of Edinburgh, Blackford Hill, Edinburgh EH9 3HJ, UK}
\affil{$^{13}$ School of Arts \& Sciences, University of Tokyo 3-8-1, Komaba, Meguro, Tokyo, 153-8902, Japan}
\affil{$^{14}$ Department of Physics, Nagoya University, Furo-cho, Chikusa, Nagoya, Aichi, 464-8602, Japan}
\affil{$^{15}$ Max-Planck-Institut für Astronomie, Königstuhl 17, 69117 Heidelberg, Germany}
\affil{$^{16}$ Leiden Observatory, Leiden University, P.O. Box 9513, 2300 RA, Leiden, The Netherlands}
}%

\jid{PASA}
\doi{10.1017/pas.\the\year.xxx}
\jyear{\the\year}

\usepackage[authoryear]{natbib}
\bibpunct{(}{)}{;}{a}{}{,}
\usepackage{aas_macros}
\usepackage{pifont}
\usepackage[]{hyperref} 
\usepackage[table]{xcolor}
\hypersetup{colorlinks,citecolor=blue,linkcolor=blue,urlcolor=blue}
\usepackage{multirow}

\newcommand{\tickmark}{\ding{51}}%
\newcommand{\crossmark}{\ding{55}}%

\usepackage{array}
\newcolumntype{L}[1]{>{\raggedright\let\newline\\\arraybackslash\hspace{0pt}}m{#1}}
\newcolumntype{C}[1]{>{\centering\let\newline\\\arraybackslash\hspace{0pt}}m{#1}}
\newcolumntype{R}[1]{>{\raggedleft\let\newline\\\arraybackslash\hspace{0pt}}m{#1}}

\begin{document}%
\begin{abstract}

The Protoplanetary Discussions conference --- held in Edinburgh, UK, from 7$^{\rm th}$--11$^{\rm th}$ March 2016 ---  included several open sessions led by participants. This paper reports on the discussions collectively concerned with the multiphysics modelling of protoplanetary discs, including the self-consistent calculation of gas and dust dynamics, radiative transfer and chemistry.  After a short introduction to each of these disciplines in isolation, we identify a series of burning questions and \emph{grand challenges} associated with their continuing development and integration.  We then discuss potential pathways towards solving these challenges, grouped by strategical, technical and collaborative developments. This paper is not intended to be a review, but rather to motivate and direct future research and collaboration across typically distinct fields based on \textit{community driven input}, to encourage further progress in our understanding of circumstellar and protoplanetary discs. 


%

\end{abstract}
\begin{keywords}
Protoplanetary discs ---  Planetary Systems: Formation ---  Chemistry --- Dust --- \\ Radiative Transfer --- Hydrodynamics
\end{keywords}
\maketitle%
\section{INTRODUCTION}
\label{sec:intro}
 For the first time in history, spatially resolved observations of the structures \emph{within} protoplanetary discs are being obtained  (see review by \citealt{2016PASA...33...13C}). This has revealed a wealth of sub-structure, including rings and gaps \citep{2015ApJ...808L...3A,2016ApJ...820L..40A,2016MNRAS.458L..29C}, spirals \citep[e.g.][]{2013A&A...560A.105G,2015A&A...578L...6B,2015ApJ...813L...2W}, warps \citep[e.g.][]{2015ApJ...811...92C}, shadows \citep[e.g.][]{2016arXiv160300481S}, cavities \citep[e.g.][]{2011ApJ...732...42A} and dust traps \citep[e.g.][]{2013Sci...340.1199V,2016A&A...585A..58V}. These recent observations, combined with the huge diversity of exoplanetary systems discovered over recent years \citep{2015ARA&A..53..409W}, has stimulated a new wave of rapid development in the modelling of protoplanetary discs, to better understand their evolution, along with their connection to the planet formation process \citep[e.g.][]{2006RPPh...69..119P}. 
 
 
 \smallskip

Understanding the evolution of discs, the structures that we are observing within them and the planet formation process presents a substantial challenge to modellers.  Discs are composed of non-primordial material spanning conditions ranging from cold, extremely dense and molecular, through to diffuse, hot and ionised. Densities and temperatures vary by $\sim$ 10 and 3 orders of magnitude, respectively. The basic chemical composition of discs alone is the subject of at least four complex research fields distinguished by the local matter conditions and radiation field: dust grains, gas-grain chemistry, photon dominated chemistry and photoionisation \citep[e.g.][]{2009ApJ...690.1539G, 2015A&A...574A.138T, 2015A&A...582A..88W, 2015ApJ...804...29G}.  The situation is even more challenging since the observational determination of a disc's composition is often degenerate, making direct comparison between observations and theory (and thus validation of our models) difficult \citep[e.g.][]{2008A&A...492..451M, 2016A&A...586A.103W,2016arXiv160507780M,2016MNRAS.461..385B,2016arXiv160505093K}.  

\smallskip

\begin{figure*}
\includegraphics[width=\textwidth]{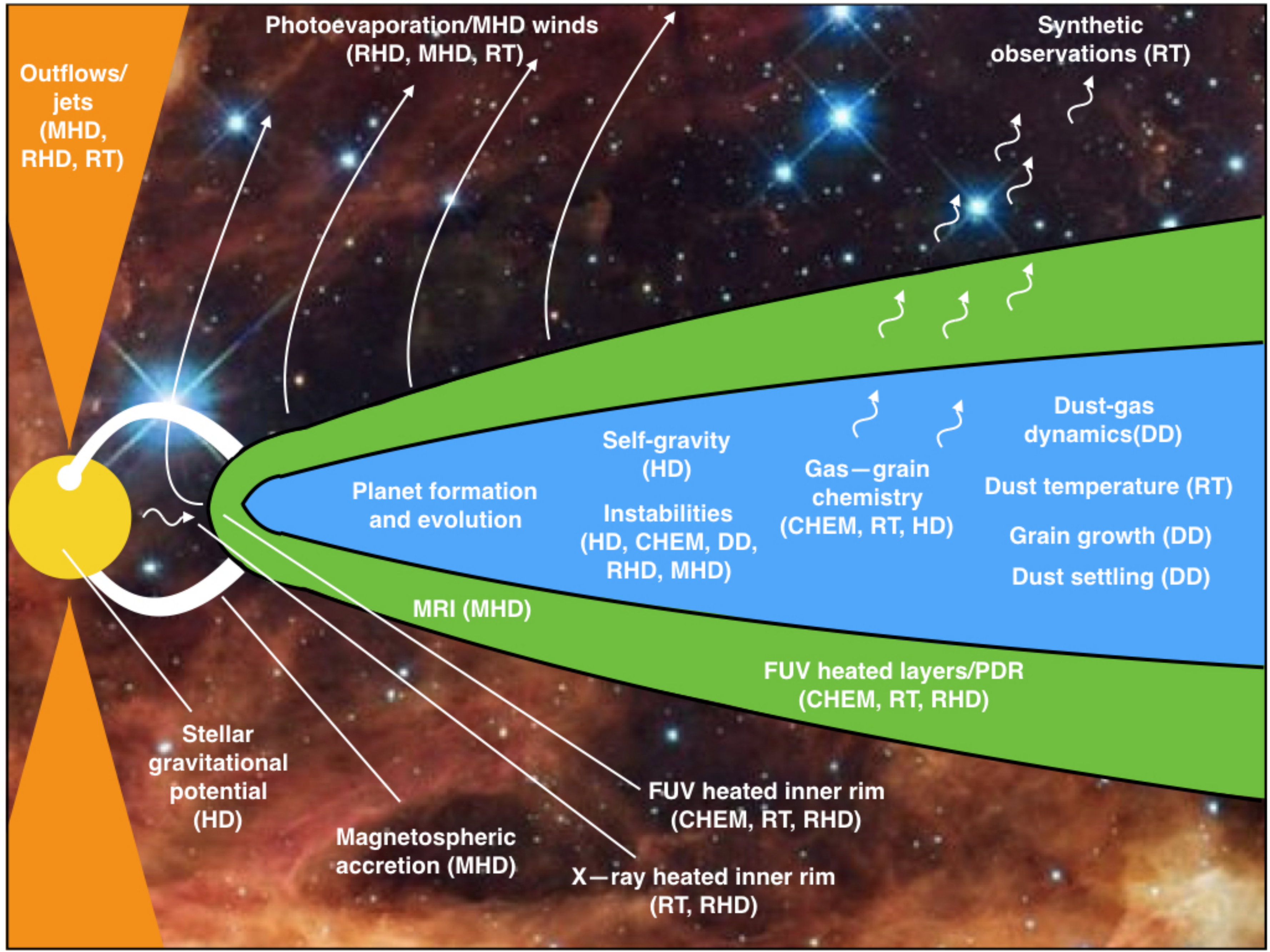}
\caption{A protoplanetary disc schematic highlighting some of the key disc mechanisms and physics we are required to model to capture them (in parentheses). These physical ingredients are hydrodynamics (HD),  magnetohydrodynamics (MHD), radiation hydrodynamics (RHD), radiative transfer (RT), chemistry (CHEM) and dust dynamics (DD). The background image is a subset of a Hubble observation of R136, credit: NASA, ESA, and F. Paresce (INAF-IASF, Bologna, Italy), R. O'Connell (University of Virginia, Charlottesville), and the Wide Field Camera 3 Science Oversight Committee.}
\label{fig:processes}
\end{figure*}

The dynamics of protoplanetary discs are also extremely challenging. The gravitational potential from the parent star, self-gravity of the disc, hydrodynamic torques in the disc, radiation from the parent star or other nearby stars, dust, and (non-ideal) magnetohydrodynamics all play important roles \citep{bodenheimer_1995, dullemond_2007, lodato_2008, 2011ARA&A..49..195A, armitage_2015}. Furthermore, the dynamical evolution of dust grains with moderate Stokes numbers $St \gtrsim 0.01$ must be solved in addition to the gas dynamics \citep[for a recent review, see][]{2014prpl.conf..339T}. Discs are also not necessarily in a steady state, and can be subject to a range of instabilities, such as gravitational fragmentation \citep{2007prpl.conf..607D, 2015MNRAS.451.3987Y, 2015MNRAS.447..836F, 2015MNRAS.454.2529M, 2016MNRAS.458.3597T}, the streaming instability \citep{2005ApJ...620..459Y}, Rossby wave instability \citep[e.g.][]{1999ApJ...513..805L, 2001A&A...380..750T, 2008A&A...491L..41L, 2009A&A...493.1125L},  baroclinic and vertical shear instabilities, which can form and grow vortex structures \citep{lyra_2011,lesur_2010,2013MNRAS.435.2610N,richard_2016}, the magneto-rotational instability \citep[e.g.][]{balbus_1991, 2003RMxAC..18...92R} and dust-settling induced vortices \citep{2015MNRAS.453L..78L, 2016MNRAS.457L..54L}.
The local environment can also significantly modify disc evolution via mass transfer from the ambient medium onto the disc \citep{2015A&A...573A...5V, 2015MNRAS.449..662L},  nearby radiation sources \citep[e.g.][]{2000AJ....119.2919B, 2002ApJ...566..315H, 2003ApJ...587L.105S, 2004ApJ...611..360A, 2012ApJ...746L..21W, 2016MNRAS.457.3593F} and tidal encounters \citep[e.g.][]{1993MNRAS.261..190C, 2012A&A...546L...1D, 2014MNRAS.441.2094R, 2015A&A...577A.115V, 2015MNRAS.449.1996D, 2016arXiv160607431V}. A summary of some of the key processes (local, not environmental) that modellers attempt to capture in discs is given in Figure \ref{fig:processes}. 

\smallskip

This physically	rich environment is made even more complex given that most of these dynamic, magnetic, radiative and chemical processes are interlinked. For example, the effect of magnetic fields depend upon the ion density, which in turn is determined by the composition, which in turn depends upon the radiation field (e.g. due to photoionisation of atoms, photodissociation of molecules and determination of the thermal properties through processes such as line and continuum cooling).  Another distinct coupling is the interaction between the gravitational instability and the magnetorotational instability, which has been well-studied in the disc community using semi-analytic models as the cause of an accretion limit cycle causing protostellar outburst phenomena \citep{armitage_2001}, but is only now being investigated with self-consistent hydrodynamic simulations \citep[e.g.][]{bae_2014}.  Another example is that the radiation field in a disc is sensitive to the distribution of small dust grains \citep[the motions of which may also be influenced by the radiation field, e.g.][]{2016MNRAS.461..742H} which in turn is sensitive to dynamical effects such as shadowing caused by warping of the inner disc \citep{2015ApJ...798L..44M,2016arXiv160300481S}. Furthermore, radiative heating increases the gas sound speed, and hence the amount of turbulent motion transferred to dust grains via gas-dust coupling, which influences grain-grain collisions and therefore the growth and fragmentation of dust \citep[e.g.][]{2014prpl.conf..339T}. As a final example, gravitational instability and fragmentation in discs is sensitive to radiation \citep[e.g.][]{2010MNRAS.406.2279M, 2013MNRAS.430.2082F} and magnetic fields \citep{2007MNRAS.377...77P,wurster_2016}, and can induce dramatic effects in the chemical composition of discs \citep[see section  \ref{fig:chem-hydro},][]{2011MNRAS.417.2950I, 2015MNRAS.453.1147E}.

\smallskip

Given the importance of these links, ultimately one wishes to identify which physical processes affect each other in a non-negligible fashion, and to model all of them simultaneously. The modelling of protoplanetary discs is therefore a daunting task --- what might be termed a \emph{grand challenge}. Each physical mechanism requires sufficient rigour and detail that modelling them constitutes an active field of protoplanetary disc research in their own right \citep[for reviews of physical processes in protoplanetary discs, see e.g.][]{1998apsf.book.....H, 2011ARA&A..49..195A, 2011ARA&A..49...67W, armitage_2015}. In practice, we have neither the numerical tools nor computational resources to achieve such multiphysics modelling of protoplanetary discs at present (nor in the immediate future). However, we can set out a roadmap towards this goal while outlining the more achievable milestones along the way.

\smallskip

In this paper, motivated by group discussion sessions at the ``Protoplanetary Discussions'' conference in Edinburgh\footnote{\url{http://www-star.st-and.ac.uk/ppdiscs/}}, we ultimately aim to stimulate progress in the multiphysics modelling of protoplanetary discs in order to deepen our understanding of them. This paper is presented in parallel with a second paper which focuses on the observations required to advance our understanding of discs \citep{siciliaAguilarInPrep}. Although our focus here is new numerical methods and the questions they might answer, it is important to remember that there are still many unsolved problems that can be tackled with existing techniques. Additionally, new numerical methods are likely to be computationally expensive so there will be many problems that are \textit{better} tackled using existing techniques \citep[e.g. parametric models used to interpret observations][]{2014ApJ...788...59W}.  Furthermore this paper is not exhaustive, there will certainly be fruitful avenues of theoretical research into protoplanetary discs that are not discussed here (in particular regarding magnetic fields and the details of planet formation itself).

\smallskip

  The structure of this paper is as follows - in Section \ref{sec:overview} we provide an overview of some core ingredients of disc modelling. In Section \ref{sec:challenges} we then present a series of mid and long term challenges to motivate future development. Finally in Sections \ref{Strategy}--\ref{Collaboration} we discuss  pathways towards meeting the challenges in terms of strategical, technical and collaborative developments. 

\bigskip

\begin{figure*}
\includegraphics[width=\textwidth]{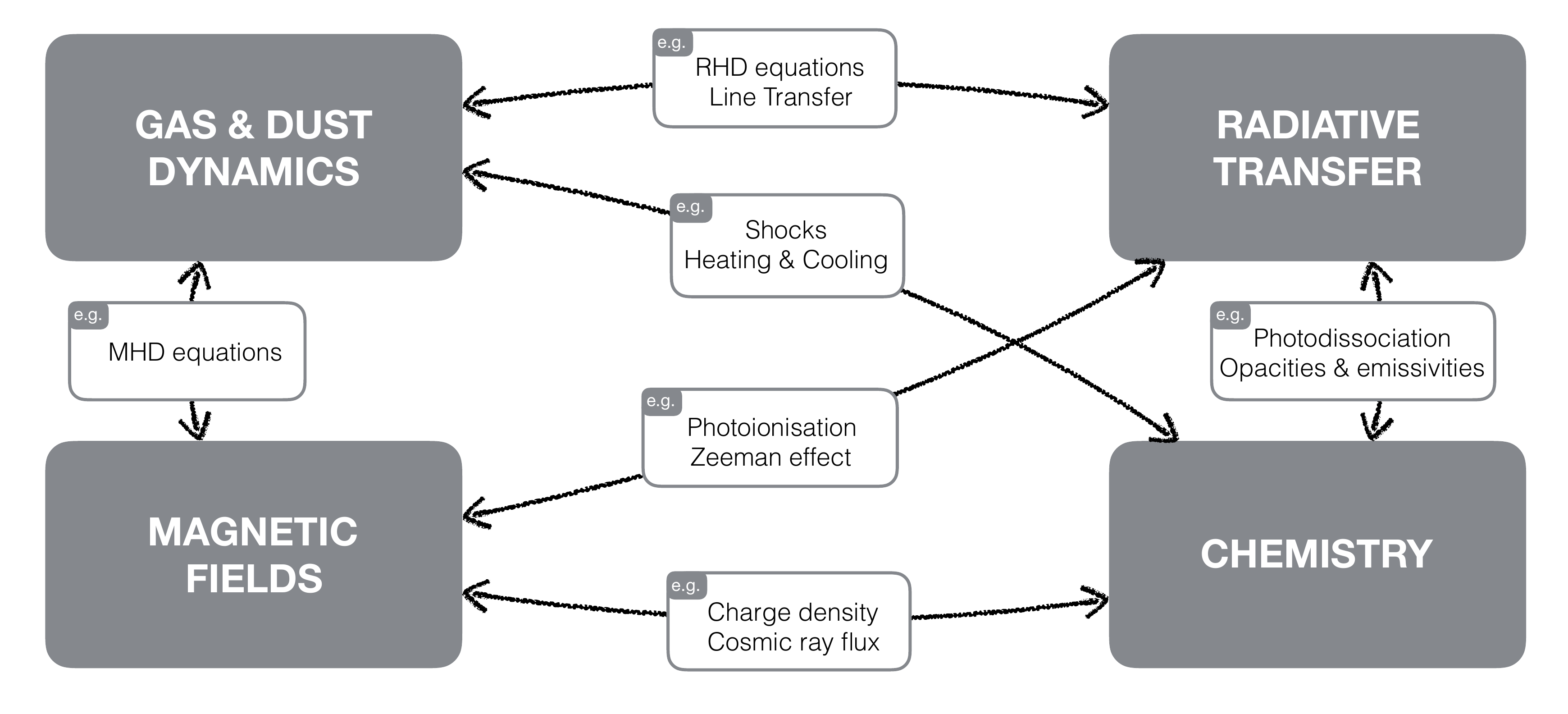}
\caption{An illustration of the core disciplines in protoplanetary disc modelling: gas \& dust dynamics, magnetic fields, radiative transfer, and chemistry.  Each discipline is a field in its own right, subject to intensive study.  However, they are all closely interlinked, affecting each other in a number of ways, of which we illustrate a few representative examples.  It is this interdependence between fields that necessitates the drive towards multiphysics modelling of  protoplanetary discs.}
\label{fig:connections}
\end{figure*}

\section{AN OVERVIEW OF CURRENT TECHNIQUES}
\label{sec:overview}
We begin by providing a overview of some of the core ingredients of protoplanetary disc modelling, to introduce concepts and provide context for the rest of the paper. This is by no means intended to be a comprehensive review, rather it should provide some basic platform from which a reader unfamiliar with certain concepts can proceed through the rest of the paper. Figure \ref{fig:connections} illustrates the four core disciplines that comprise the majority of protoplanetary disc modelling: gas and dust dynamics, magnetic fields, radiative transfer and chemistry. As shown, these topics are all fundamentally linked. It is this interdependence that raises the possibility that multiphysics modelling will be important and is hence a key focus of this paper.

\subsection{(Magneto-) Hydrodynamics}
Solving for the motion of fluids as a function of time is a key ingredient for understanding the evolution of protoplanetary discs.  Numerical hydrodynamics is a relatively mature field. Numerical solvers are either Eulerian or Lagrangian in character. Eulerian solvers trace flows across fixed discrete spatial elements, while Lagrangian solvers follow the motion of the flow. In protostellar disc simulations, the majority of hydro solvers are either Eulerian/Lagrangian grid based simulators, or the fully Lagrangian Smoothed Particle Hydrodynamics.

Depending on the resolution requirements, solvers are either \emph{global}, in that the entire disc extent is simulated together, or \emph{local}, where a region in the disc is simulated at high resolution, with appropriate boundary conditions to reflect the surrounding disc environment. Which construction is best used is dependent upon the problem being studied, as we discuss below. 

\bigskip

\subsubsection{Global disc simulations}
Historically, the primary challenge for global simulations of protoplanetary discs with Eulerian codes was the Keplerian flow --- advection of material at supersonic speeds across a stationary mesh is a recipe for high numerical diffusion. This has now been overcome with, for example, the {\sc fargo} algorithm \citep{2000A&AS..141..165M}, implemented in both the {\sc fargo} \citep{2000A&AS..141..165M,2008ApJ...678..483B,2016ApJS..223...11B} and {\sc pluto} \citep{2007ApJS..170..228M} codes. Eulerian codes perform best when the flow is aligned with the grid. This means that cylindrical or spherical grids are preferable which, when applicable, offer the best accuracy currently possible of any technique for a given level of computational expense or resolution. However, this means that adaptive mesh refinement \citep{berger_1989}, which is mainly (but not exclusively) developed for Cartesian meshes, is not typically used \citep[an example exception is][]{2004A&A...425L...9P}. Furthermore, simulating warped, twisted or broken discs remains difficult (e.g.\ \citealt{2010A&A...511A..77F}).  


Lagrangian schemes such as smoothed particle hydrodynamics \citep[SPH, for reviews see e.g.][]{1992ARA&A..30..543M,2012JCoPh.231..759P} are well suited to more geometrically complex global disc simulations because advection is computed exactly, angular momentum can be exactly conserved (e.g. an orbit can be correctly simulated with one particle) and there is no preferred geometry. Numerical propagation of warps using SPH has been shown to closely match the predictions by \citet{ogilvie99} of $\alpha$-disc theory \citep{2010MNRAS.405.1212L}. In particular, a generic outcome of discs that are misaligned with respect to the orbits of central binaries or companions is that the disc `tears' \citep{2012ApJ...757L..24N,2013MNRAS.434.1946N,2015MNRAS.448.1526N} or breaks \citep{2012MNRAS.421.1201N,2013MNRAS.433.2142F,2015MNRAS.449.1251D}.  Such behaviour is well modelled by SPH codes, and appears to be relevant to observed protoplanetary discs, including HK Tau \citep{1998ApJ...502L..65S}, KH15D \citep{2013MNRAS.433.2157L}, and HD142527 \citep{2015ApJ...811...92C}. A limitation of the SPH approach is that the particles adaptively trace the densest regions, low density components of the disc, e.g. gaps and the disc upper layers, can therefore be under-resolved \citep[e.g.][]{2006MNRAS.370..529D}.

\subsubsection{Local simulations}
 The most common technique utilised for local simulations of discs is the Cartesian shearing box \citep{hgb95,guangammie08}. This imposes the shear flow in a subset of a disc and allows for high resolution simulations of disc microphysics in a Cartesian geometry, well suited to most Eulerian codes. This means that all the sophistication of modern Godunov-based hydrodynamics can be applied \citep[there are many textbooks covering grid based hydrodynamics, e.g.][]{toro2013riemann}. This approach has been used almost exclusively for simulating the magnetorotational \citep[see][and references within]{balbus_2003} and other instabilities --- in particular the streaming instability \citep[e.g.][]{2005ApJ...620..459Y,2007ApJ...662..613Y,2007Natur.448.1022J,2010ApJ...722.1437B} --- in discs. Though other applications include the study of magnetically driven disc winds \citep[e.g.][]{2009ApJ...691L..49S, 2010ApJ...718.1289S}. 
 
 By contrast, at present there is no particular advantage to using Lagrangian schemes for local disc simulations. The cost for comparable results in cartesian boxes is up to an order of magnitude higher in SPH compared to Eulerian codes \citep[e.g.][]{2008MNRAS.390.1267T,2010MNRAS.406.1659P}, mainly due to the additional costs associated with finding neighbouring particles, and the algorithms tend to be more dissipative than their grid-based counterparts, particularly when the flow is well matched to the grid geometry. However, Lagrangian techniques can accommodae open boundary conditions more naturally, so may offer advantages for certain problems in the future.

\subsubsection{Other codes}
In recent years several new hydrodynamic solver methods have appeared.  This broad class of Arbitrary Lagrangian Eulerian methods (ALE) offer the user the ability to switch between Lagrangian and Eulerian formalisms smoothly, in some cases during simulation runtime.  Such ALE solvers include moving mesh codes \citep{Arepo,2011arXiv1109.2218S,2011ApJS..197...15D} and meshless codes \citep{2012ApJS..200....6M,2012ApJS..200....7M,gizmo}. This extreme flexibility in approach appears to offer highly conservative schemes and adaptive resolution while capturing mixing and shear instabilities with high fidelity.  The relative youth of these techniques (at least, in their application to computational astrophysics) means the full extent of weaknesses and strengths in these approaches remains to be seen (e.g. the ``grid noise'' encountered during mesh regularisation; \citealt{2015MNRAS.452.3853M}) although early applications to protostellar discs appear to be promising \citep[see e.g.][]{2014MNRAS.445.3475M}.

Another recent development in numerical astrophysical fluid dynamics is the use of discontinuous Galerkin methods (which have a long history of application in the mathematical community). These grid based techniques offer accurate, high order solutions in a manner that is readily applied to adaptive meshes, and that scale efficiently on modern high performance computing facilities. In the astrophysical community, discontinuous Galerkin algorithms have now been implemented in both Cartesian \citep[e.g. the \textsc{tenet} code; ][]{2015MNRAS.453.4278S} and moving Voronoi mesh \citep[e.g. the \textsc{arepo} code; ][]{2014MNRAS.437..397M} frameworks.

\subsubsection{Magnetic fields}


The above hydrodynamic solvers are able to include the evolution of the magnetic field in their fundamental equations. This has been most easily incorporated in Eulerian solvers, with mature magnetohydrodynamic (MHD) implementations in, for example, the \textsc{athena} \citep{2008ApJS..178..137S}, \textsc{enzo} \citep[][]{enzo}, {\sc fargo} \citep{2016ApJS..223...11B}, {\sc pluto} \citep{2007ApJS..170..228M} and \textsc{pencil} \citep{2002CoPhC.147..471B} codes. SPH and other meshless codes can now also incorporate MHD (see review by \citealt{2012JCoPh.231..759P}), provided that the $\nabla \cdot\mathbf{B}=0$ condition can be sustained, for example using divergence cleaning techniques \citep{tricco_2012}. Note however, that MHD with SPH is not a mature approach and is therefore somewhat less robust than Eulerian MHD at present \citep[e.g.][]{2016arXiv160606972L}.

While ideal MHD disc simulations have been conducted for some time \citep[see][and references within]{balbus_2003}, particularly important for protostellar discs is the role of non-ideal MHD, ever since the idea of a `dead zone' was proposed by \citet{1996ApJ...457..355G}. More recently, the interplay between the Hall effect, ambipolar diffusion and Ohmic diffusion is yielding new turbulent behaviour \citep{2002ApJ...570..314S,simon_2015}, new forms of instability and zonal flows in both MRI-active and `dead zone' regions \citep[e.g][]{kunz_2013,bai_2014}, not to mention addressing the so-called magnetic braking catastrophe that suppresses disc formation in ideal MHD \citep{2015ApJ...810L..26T,wurster_2016} (see recent review by \citealt{2016PASA...33...10T}, this volume).

For more general modelling of young stellar systems, global simulations are particularly important for modelling the launching of magnetised winds from the star and/or disc, and jets from the central star \citep[e.g.][]{casse_2007,Bai_2014wind,lovelace_2014,2014ApJ...784..121S, staff_2016}.


\subsubsection{Remarks on hydrodynamics}
In summary, there are a number of options available as to how to model the (magneto-)hydrodynamical evolution of a disc - the problem one is addressing determines which method is most appropriate. This ``horses for courses'' approach is important, and is likely to extend to efforts which hope to further include elements from the other disciplines of disc modelling such as chemistry and radiation transport.

\subsection{Dust-gas dynamics}
\label{sec:dustdyn}
The dynamics of small dust grains (Stokes number $\ll1$) is typically well coupled to that of the gas. For larger grains, however, the dust and gas dynamics can be decoupled. Properly modelling these decoupled motions is important both for disc dynamics, but also for interpreting observations.  This latter point is particularly prudent given that some of the most important disc observations in recent years are millimetre continuum observations (i.e.\ of dust). For example, decoupled dust and gas dynamics is apparently important for understanding the symmetric gaps observed in discs \citep[e.g.][]{2015MNRAS.453L..73D, 2016ApJ...818...76J, 2016MNRAS.459.2790R}. 

Approaches for modelling the dynamics of dust grains that are decoupled from the motions of the gas are often distinguished by whether they use a single or two fluid approach, both of which we discuss below.

\subsubsection{Two fluid or `hybrid' schemes}
In an SPH framework, the two-fluid approach sees the dust and gas as separate particle populations, the dynamics for which are solved separately \citep{1995CoPhC..87..225M,2005A&A...443..185B,2012MNRAS.420.2345L,2012MNRAS.420.2365L,2014MNRAS.443..927L, 2015MNRAS.452.3932B}. In grid-based methods the dust is typically simulated as a particle population, with the hydrodynamics computed on the grid \citep[e.g.][]{2007A&A...462..355P, 2008A&A...479..883L, 2010JCoPh.229.3916M,2010ApJS..190..297B,2015A&A...574A..68F, 2016MNRAS.458.3927B, 2016ApJS..224...39Y} --- hence usually referred to as a `hybrid' approach. The `hybrid' or `two fluid' approaches are best suited to decoupled grains with Stokes number $\gtrsim 1$, where the interaction can be computed explicitly.

The traditional difficulty when dust is modelled by a separate set of particles is that short timesteps are required for small grains (Stokes numbers $\ll 1$), requiring implicit timestepping schemes \citep{1997JCoPh.138..801M,2010ApJS..190..297B,2010JCoPh.229.3916M,2012MNRAS.420.2365L}. However, \citet{2012MNRAS.420.2345L} showed that simulating tightly coupled grains this way leads to `overdamping' of the mixture, becoming increasingly inaccurate for small Stokes numbers, caused by the need to spatially resolve the `stopping length' $l \sim c_{\rm s} t_{\rm s}$ (where $c_{\rm s}$ is the sound speed and $t_{\rm s}$ is the stopping time). A similar issue was noted by \citet{2010JCoPh.229.3916M} in the context of grid based codes, finding only first order convergence in the `stiff' regime when the stopping time is shorter than the Courant timestep. However, by making use of the analytical solutions for the motion under drag forces that respect the underlying problem this dissipation can be substantially reduced \citep[or entirely avoided in the limit of negligible dust mass,][]{2014MNRAS.443..927L}.

\subsubsection{Single fluid schemes}
In the single fluid approach the dust parameters (dust to gas ratio, relative velocity) are properties of the `mixture'. In SPH this means that a single population of SPH particles is used, representing the total fluid mass, with dust properties updated on each `mixture' particle \citep{2014MNRAS.440.2136L, 2014MNRAS.440.2147L, 2014MNRAS.444.1940L,2015MNRAS.451..813P,2016MNRAS.461..742H}. The same approach on a grid means evolving the dust density on the grid (called a `two fluid' approach by \citealt{2010JCoPh.229.3916M} --- though not to be confused with the two fluid approach mentioned above --- to distinguish it from the `hybrid' grid-plus-particles method). This is sometimes achieved using the approach suggested by \citet{2005ApJ...634.1353J} based on the `short friction time' or `terminal velocity approximation' for small grains. Here the dust continuity equation is solved and the dust velocity is set equal to the gas velocity plus the stopping time times the differential forces between the gas and dust mixture. This is similar to the `diffusion approximation for dust' derived by \citet{2014MNRAS.440.2136L} and implemented in SPH by \citet{2015MNRAS.451..813P} with an important caveat -- that this formulation is only valid when the dust fraction is small (since it assumes that the gas velocity equals the barycentric velocity of the mixture). This assumption can easily be relaxed, at no additional computational expense, as shown by \citep{2014MNRAS.440.2136L}.

An attraction of fluid based dust models is that within their domain of validity they provide a high degree of accuracy for their computational cost, while particle approaches typically suffer from sampling noise. However, the fluid approach is equivalent to using a moment based method for solving the radiative transfer equations (see Section \ref{sec:rt}) where all moments of order greater than unity (or even zero in the short-friction time approach) are discarded. This means that in cases where the dust velocity becomes multi-valued the result may converge to the wrong answer. Possible examples of when this can occur include settling (for $S_t > 1$), turbulent motion ($S_t \gtrsim R_e^{-1/2} \sim 10^{-4}$ in astrophysical flows, although Reynolds nymbers, $R_e \gtrsim 10^3$, are rarely achieved numerically \citealt{2002Natur.419..151F,2007A&A...466..413O}), strong gravitational scattering, and in convergent flows at curved shocks. By including higher order moments, the fluid approximation could be extended to support multi-valued flows and thus support both large and small grains \citep{Chalons2010a,Chalons2010b,Yuan2011,Yuan2012}.

\subsubsection{Dust post-processing approaches}

While the dynamical evolution of discs is clearly of importance to many problems, there are many cases in which the dynamic time-scales are very different to other processes (see also Section \ref{sec:ResTime} of this paper). For example, the short radiative time-scale in discs has led to the standard approach of treating them as isothermal. Similarly, since dust growth often occurs on much longer timescales ($ > 10^4\,{\rm yr}$) the approach of post-processing the dust evolution according to some average over the short term dynamics can be viable. For example \cite{2008A&A...480..859B} and \cite{2010A&A...513A..79B} evolve the gas disc until a steady state is reached and then evolve the dust against this steady gas background.

This approach has also been applied to transition discs and discs with massive planets embedded, in particular following the growth of large particles trapped inside pressure maxima \citep{2015A&A...573A...9P,2016A&A...585A..35P}. Similarly, \citet{2015MNRAS.451..974D} applied this approach to self-gravitating discs in order to predict scattered light images. \cite{2016ApJ...821....3M} have also studied the motions of dust grains against a fixed gas background for the scenario of magneto-rotationally driven winds.  However, we note that this approach can be fraught with difficulty, since it is difficult to know a priori what the representative average of the disc should be within which to evolve the dust. For example, particles with $St \sim 1$ can become trapped in the spiral arms of self-gravitating discs (or other pressure maxima), making azimuthal averaging unreliable. Similarly, although \citet{2016MNRAS.459.2790R} showed that azimuthal averaging works well for transition discs formed by planets of order a Jupiter mass or less, ignoring the gas-dynamics completely would predict an incorrect surface density profile and thus also incorrect growth rates. However, when the effects of combined dust-gas dynamics are taken properly into account (e.g. the short-friction time approximation can be used with hydrodynamic models to predict the evolution of dust grains $1\,{\rm mm}$ or smaller in transition discs), the post-processing approach will undoubtedly continue to provide important insights. 

Conversely, coupling to live simulations of the dust/gas dynamics may prove to be essential for understanding some phenomena. For example, \citet{2015MNRAS.454L..36G} showed that by incorporating grain growth, radial drift and feedback that self-induced dust traps may arise \citep[to be explored in more detail in][]{Gonzalez2016InPrep}. There will be many other important cases that likely require live simulations, for example, understanding whether planet formation can occur via the streaming instability in dust traps will require models that can show whether grains can grow to the required sizes without destabilizing the trap \citep[e.g.][]{2012ApJ...747...11K,2016A&A...591A..86T}.

\subsubsection{Remarks on dust dynamics}

To date there are virtually no simulations where both small and large grains are \textit{directly} simultaneously evolved alongside the gas, in 3D, including the backreaction on the gas \citep[though considerable progress towards this has been made by][]{2007A&A...462..355P, 2008A&A...479..883L, 2015P&SS..116...48G, 2015MNRAS.454L..36G}. Such a combination is important, because the grains, particularly when the dust-to-gas ratio becomes high, exert a backreaction on the gas, which in turn modifies the dynamics of the other grain species. For example, \citet{2014MNRAS.444.1940L} showed that under certain conditions effects from the dynamics of multiple grain species could lead to the outward rather than inward migration of pebble-sized grains in discs. While the large grain populations with $S_{\rm t} \gtrsim 1$ are more interesting dynamically because they are more decoupled from the gas, modelling the small grains is necessary for coupling with radiative transfer and thus for comparison with observations. \cite{2007A&A...462..355P} and \cite{2008A&A...479..883L} do model a distribution of grain sizes using a particle appraoch, but not in regimes where the backreaction on to the gas is accounted for. Another often used approach is to perform a series of single grain-size simulations, and merge the results \citep[e.g.][]{2012A&A...547A..58G, 2015MNRAS.453L..73D}. While these approaches neglect any feedback that the grain species have on the gas dynamics, they have proved a useful tool for direct comparison with observations.
 
From the perspective of dust dynamics, a long term goal would be to model the dynamics of the whole grain population in discs simultaneously, in 3D, including the effects of the dust on the gas dynamics. Some progress towards this was made by  \citet{2010ApJS..190..297B, 2014MNRAS.444.1940L}, showing how multiple grain species can be treated simultaneously within a one-fluid approach, but this is not yet implemented in any numerical code. Modelling the entire grain population would open the possibility of coupling the dust dynamics directly to the radiative transfer. In turn, the radiative transfer could then be used to set the gas temperature profile in the disc, allowing for thermodynamic feedback between the grain dynamics and the gas and the coupling to chemistry.

\subsection{Radiative transfer}
\label{sec:rt}
The transport of radiation through matter is important for three primary reasons. Firstly, radiation can modify the composition and thermal properties of matter. For example, changing the composition and heating through mechanisms such as photoionisation and photodissociation and cooling it through the escape of line emission. Radiation can also set the dust temperature, which is determined by radiative equilibrium between thermal emission from the grains and the local radiation field \citep[there are a number of textbooks with extensive discussion of these topics, such as][]{1978ppim.book.....S, 1979rpa..book.....R, 2006agna.book.....O}. This impact on the composition and thermal structure drives many macroscopic processes in discs (see e.g. section \ref{sec:intro}, Figure \ref{fig:processes}). Secondly, radiation pressure can directly impart a force upon matter, altering the dynamics. Finally, radiation is what is \textit {actually} observed. Radiative transfer is therefore required to make the most meaningful and robust comparisons between theoretical models and observations. 

Since radiative transfer is fundamentally coupled to matter (influencing the composition and temperature, which in turn modifies opacities and emissivities), the coupling of radiation transport and chemistry is already an established field, which will be discussed further in section \ref{sec:chem}. 

For purely dynamical applications the only quantities of interest from radiative transfer are a temperature/pressure estimate and/or a radiation pressure estimate. To this end, popular techniques are flux limited diffusion (FLD) and similar moment methods, owing to their relatively minimal computational expense compared with more detailed radiative transfer methods \citep[e.g.][]{1981ApJ...248..321L, 2004MNRAS.353.1078W, whitehouse_2005}. In FLD schemes, the directional properties of the radiation field are replaced by angle averaged ones and the radiative transfer problem is solved using a diffusion equation.  
FLD has long been applied in optically thick regimes without sharp density contrasts, but can generate spurious results where this is not the case \citep{2014ASSP...36..127O, 2013A&A...555A...7K}.  Most modern applications of FLD account for this failure at low optical depth by using boundary conditions \citep[e.g.][]{mayer_2007}, or using hybrid methods to allow the system to radiate energy away from optically thin regions \citep[e.g.][]{2007ApJ...665.1254B,forgan_2009}. Other approximate temperature prescriptions have also been developed that are tailored to model the effect of higher energy extreme ultraviolet (EUV) and X-ray photons from the parent star on the disc evolution \citep[e.g.][]{2006MNRAS.369..216A, 2006MNRAS.369..229A, 2010MNRAS.401.1415O,2011MNRAS.412...13O, 2012MNRAS.422.1880O, 2016MNRAS.457.1905H}.

More rigorous radiation transport methods have historically typically been confined to computing synthetic observables, where the density structure is based on snapshots from dynamical models, hydrostatic equilibrium in a simple disc, or a parametric model. Perhaps the most popular method in this context is Monte Carlo radiative transfer \citep{1999A&A...344..282L}, which is used by the well known codes \textsc{radmc-3d} \citep{2012ascl.soft02015D}, \textsc{mcmax} \citep{2009A&A...497..155M}, \textsc{hyperion} \citep{2011A&A...536A..79R}, \textsc{mcfost} \citep{2006A&A...459..797P} and \textsc{torus} \citep[][also discussed below]{2015MNRAS.448.3156H}. Monte Carlo radiation transport typically involves  breaking the energy from radiative sources into discrete packets, which are propagated through space in a random walk akin to the propagation of real photons through matter (e.g. including scattering and absorption/re-emission events). This provides an estimate of the mean intensity everywhere which can be used, for example, to solve for the ionisation state of a gas, the dust radiative equilbrium temperature, or to generate synthetic observations. The Monte Carlo approach naturally accounts for the processed radiation field (scatterings, recombination photons), works in arbitrarily geometrically complex media and also treats multi-frequency radiation transport (conversely FLD approaches typically assume that the opacity is frequency independent). 

In addition to the Monte Carlo approach, other well known methods are also the pure \citep[e.g.][]{2002MNRAS.330L..53A} and short characteristic \citep[e.g.][]{2012ApJS..199....9D} ray tracing schemes. Recently intermediate expense hybrid-methods have been developed which combine FLD and other (e.g. ray--tracing) methods to offer a better balance between the accuracy of a more sophisticated scheme and the speed of FLD for dynamical applications \citep{2013A&A...555A...7K, 2014ASSP...36..127O, 2015A&A...574A..81R}.

\subsection{Chemistry}
\label{sec:chem}

Molecular line observations play a central role in determining both the conditions within, and kinematics of, protoplanetary discs. In particular, CO and its isotopologues are popular tracers which are relatively abundant, have a permanent dipole moment and estimates of canonical abundances in the interstellar medium (ISM). CO synthetic observations can therefore be generated relatively easily in discs by assuming the canonical abundance and that local thermodynamic equilibrium (LTE) applies, in which case the level populations are set analytically by the Boltzmann distribution \citep[e.g.][]{2014ApJ...788...59W}. However such a simple approach is not always valid. For example in discs there is evidence of departure from the canonical CO abundance \citep[e.g.][]{2013ApJ...776L..38F} and the relative abundance of isotoplogues does not necessarily scale as in the ISM \citep{2014A&A...572A..96M}. Furthermore, dust grain evolution and dynamical processes such as instabilities and planet-disc interactions can also affect the chemistry \citep[e.g.][]{2007ApJ...665.1254B, 2011MNRAS.417.2950I, 2015MNRAS.453.1147E, 2015Natur.520..198O, 2015ApJ...810..112O,  2015ApJ...807....2C, 2016ApJ...823L..18H}. Although simple CO parameterisations yield useful insights into the global properties of discs \citep[such as the disc mass, e.g.][]{2016arXiv160507780M, 2016arXiv160605646W} they are substantially more limited when it comes to probing the local properties. Given the above, more substantial chemical models will play an important role in the interpretation of modern protoplanetary disc observations. Furthermore, such models would support observations using molecules other than CO that are less easily parameterised, but could be better suited for probing certain components of a disc. In addition to interpreting observations, understanding the chemical evolution of discs will also have astrobiological implications in the connection to the chemical composition of planets themselves.

To date, almost 200 molecules have been detected in interstellar or circumstellar environments\footnote{\url{http://www.astro.uni-koeln.de/cdms/molecules}}.  The abundances of these molecules can be subject to change via a large number of chemical reactions \citep[see][for reviews]{caselli_2005, henning_2013}. In order to accurately model the evolution of even a small number of these molecules, complex computational networks of chemical reactions are needed in the form of coupled ordinary differential equations (ODEs).  Several research groups have compiled publicly-available databases of both these chemical reaction networks, and data on the rates of individual chemical reactions themselves - including the UMIST Database for Astrochemistry\footnote{\url{http://udfa.ajmarkwick.net}} \citep[UDfA;][]{millar_1997, woodall_2007, mcelroy_2013}, the Ohio State University networks\footnote{\url{http://faculty.virginia.edu/ericherb/research.html}}, and the Kinetic Database for Astrochemistry\footnote{\url{http://kida.obs.u-bordeaux1.fr}} \citep[KIDA;][]{wakelam_2012}.   Databases either contain these rates explicitly, or include how such a rate depends on local properties in the form of a parametrised expression (often via the Arrhenius-Kooij equation, \citealt{arrhenius_1889, kooij_1893}).

\begin{table} 
\begin{center}
{\small
\caption{Common gas-grain reactions in astrophysical environments.  Species are all considered to be in the gas phase, unless shown as X$_{\rm gr}$, which are considered to be located on the ice mantles of dust grains.  Photons are shown as $\gamma$ and cosmic rays are shown as $\gamma_{\rm cr}$.  Adapted from \citet{caselli_2005}.}
\label{tab:reacs}
    \begin{tabular}{l r  p{0.3cm}  l}
    \hline                                                                              
    Reaction					&		\multicolumn{3}{c}{Process} \\    
    \hline                                                                              
    Neutral-neutral			    &		A + B &$\rightarrow$& C + D				    \\ 
    Three-body                  &       A + B + M &$\rightarrow$& C + D + M           \\
    Radiative association		&		A + B &$\rightarrow$& AB + $h\nu$		        \\
    Ion-neutral                 &		A$^{+}$ + B  &$\rightarrow$& C$^{+}$ + D	    \\ 
    Dissociative recomb.        &		AB$^{+}$ + $e^{-}$ &$\rightarrow$& A + B      \\ 
    Charge transfer             &		A$^+$	+ B &$\rightarrow$& A + B$^+$	     	\\
    Photodissociation           &		AB + $\gamma$ &$\rightarrow$& A + B		        \\
    Photoionisation			    &		A + $\gamma$ &$\rightarrow$& A$^{+}$ + $e^{-}$	\\
    Cosmic-ray ionisation	    &		A + $\gamma_{\rm cr}$ &$\rightarrow$& A$^{+}$ + $e^{-}$	\\
    Adsorption                  &       A &$\rightarrow$& A$_{\rm gr}$\\    
    Desorption                  &       A$_{\rm gr}$ &$\rightarrow$& A\\    
    Grain surface               &       A$_{\rm gr}$ &$\rightarrow$& B$_{\rm gr}$\\        
\hline 
\end{tabular}
}
\end{center}
\end{table}


Chemical reactions fall into several categories and can involve a variety of reactants.  Table \ref{tab:reacs} lists the common types of astrophysical  reactions.  While the majority of reactions are concerned with gas phase species or their interaction with photons, dust grain surfaces provide a location for further chemistry to occur.  Gas phase molecules attach themselves to the surfaces of dust grains (a process known as adsorption) via two mechanisms: physisorption (involving weak van der Waals forces) or chemisorption (due to chemical valence bonds).  Once species are adsorbed, they produce layers of ices on the surface of dust grains, which allows more complex surface chemistry to occur \citep{herbst_2009}. An example of this is the process of hydrogenation, by which hydrogen reacts quickly with other surface species (including itself) to produce saturated molecules such as methane. Of particular interest for this paper is that the composition of ices on dust grains (e.g. CO-coated versus H$_2$O-coated) can also affect the subsequent evolution of the dust by affecting the sticking efficiency and coagulation and fragmentation efficiencies \citep[not discussed in detail here, but see e.g.][for further information]{ 2002ApJ...566L.121K, 2008ARA&A..46...21B, 2014prpl.conf..547J, 0004-637X-818-1-16}.  Regions that are well shielded from incident stellar radiation (such as the disc midplane) might be thought to be chemically inert, as there is not sufficient energy to overcome reaction activation barriers.  However, in such regions, ionisations caused by cosmic rays can induce ion-molecule reaction sequences that dominate much of the gas-phase chemistry, including the production of secondary cosmic-ray-induced photons.  Increased densities in the disc midplane also mean that three-body reactions in the gas phase will begin to have an important effect on the chemistry.  In these cases, a third body (M, the most abundant species, often molecular hydrogen) acts an a non-reacting catalyst. 

In addition to (closely coupled to) the computation of abundances is the computation of the temperature. This is determined by the heating and cooling rates, which are themselves set by, to name just a few: radiative processes (e.g. photoionisation heating and line cooling), dust/PAH's (e.g. PAH heating and grain radiative cooling), chemical processes (e.g. exothermic reactions), hydrodynamic work/viscous heating and cosmic rays \citep[a review is given by][]{2015EPJWC.10200011W}. Many of these heating/cooling terms are linked to the composition of the gas, requiring chemical and thermal calculations to be solved iteratively. In principle, since the heating and cooling is also set by the dust and radiation field, it might also be necessary to iterate over the (decoupled dust-gas) dynamics and radiative transfer. 

\begin{figure*}
\includegraphics[width=\textwidth]{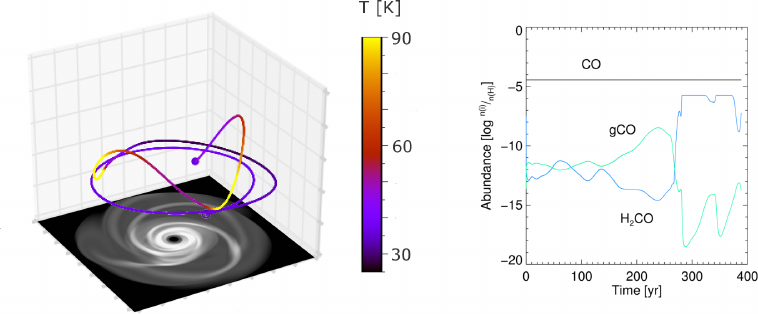}
\caption{\textit{Left}: The three-dimensional  evolution of a tracer particle in a self--gravitating disc, colour coded with temperature changes, overlaid on the final column density snapshot of the disc.  \textit{Right}: The corresponding chemical evolution of particle, showing gas-phase CO and H$_{2}$CO, and CO ice (gCO).  The shocks induced by the self--gravity of the disc have a significant impact on the chemical composition of the disc material (see \citealt[][]{2007ApJ...665.1254B, 2011MNRAS.417.2950I, 2015MNRAS.453.1147E}). }
\label{fig:chem-hydro}
\end{figure*}

Somewhat distinct from gas--grain chemistry are the photoionisation and photon dominated region (PDR) regimes, where the radiation field plays a significant role in setting the composition and temperature of a medium. Photoionised gases are composed exclusively of atoms and ions and are typically modelled more in a radiative transfer context than a chemical one. Photoionisation models are usually concerned with the transfer of EUV photons and X--rays to solve for the ionisation balance and thermal structure of a gas of assumed gas and dust abundances. Despite not requiring chemical networks, this can include a variety of processes that are not trivially captured such as resonant line transfer and inner shell ionisations  of atoms by X--rays  (the liberation of multiple electrons by a single photon).  Some examples of famous photoionisation codes are \textsc{cloudy} \citep{2013RMxAA..49..137F} and \textsc{mocassin} \citep{2003MNRAS.340.1136E}. The photoionised regime only applies to disc winds, the very surface layers/inner edge of discs and, if the disc is externally irradiated by high energy photons (e.g. from a nearby O star), components of the flow from the disc outer edge.

The PDR regime applies at the transition between photoionisation and gas--grain dominated regimes; between predominantly ionised and molecular gasses. For example in surface layers of the disc, but generally wherever matter is not optically thick to far ultraviolet (FUV) radiation. PDR modelling, like the gas--grain regime, requires a chemical network to be solved. It is also further complicated because cooling by line photons can be very important. This means that although the local radiation energy density (exciting the gas) is a single parameter, the escape probability of the line photons depends upon the extinction in all directions, i.e. it depends on the 3D structure of the surrounding space. Many PDR codes therefore compute this escape probability in one direction only, either working in 1D \citep[e.g. models such as those in][]{2007A&A...467..187R} or making some assumption about the dominant direction (e.g. vertically in the disc). Of the latter type, so called 1+1D models are particularly popular, which assume that at any given radial distance from the star the disc is in hydrostatic balance and escaping photons only consider the vertical distribution of gas at that radius \citep[e.g.][]{2009ApJ...705.1237G, 2016A&A...586A.103W}. Recently, multidimensional numerical approaches to solving PDR chemistry have appeared that do compute the 3D escape probabilities \citep{2012MNRAS.427.2100B, 2015MNRAS.454.2828B} which they do efficiently using \textsc{healpix} \citep{2005ApJ...622..759G}

\subsubsection{Remarks on chemistry and radiative transfer}

Chemical networks are used in conjunction with radiative transfer models to compute chemical abundances in various astrophysical environments. In general, the abundances are functions of temperature, density and local radiation field, though many other parameters can play a role (in particular in the regime where line cooling is important, a measure of the extinction in all directions is ideally required).  Often, the chemical networks are integrated to equilibrium in regions where the physical conditions are not thought to change significantly with time.  However, in many cases, the microphysical conditions are functions of both space \emph{and} time and are therefore not independent of dynamical processes \citep[an example of this is given in \ Figure \ref{fig:chem-hydro}, see also][]{2007ApJ...665.1254B, 2011MNRAS.417.2950I, 2015MNRAS.453.1147E, 2016arXiv160707861D}. 

Recent work has seen an increase in performing chemical evolution calculations alongside the radiative transfer calculations \citep[e.g.][]{2009ApJS..183..179B, 2009A&A...501..383W}. Furthermore, chemistry is now being coupled directly with hydrodynamic calculations: in the context of star forming regions there are the full hydrodynamic models of \cite{2010MNRAS.404....2G} and in a 1+1D disc framework there are models such as those by \cite{2009ApJ...705.1237G} which also include radiative transfer. Such coupling is particularly important in the regions of the discs where the gas is not thermally coupled to the dust (i.e.\ in the upper layers of the disc, or within the dust sublimation radius), since the gas temperature, gas abundances and level populations are strongly correlated. Unfortunately, it is in these regions of importance that 1+1D models become less applicable due to deviations from hydrostatic equilibrium \citep[for example thermally driven winds are not hydrostatic, e.g.][]{2016MNRAS.tmp..880C}. Dynamically, some chemical regimes (in particular, the PDR regime) are definitely important for understanding certain processes. For example PDR physics is required to model FUV driven photoevaporative flows from the outer edge of discs \citep{2004ApJ...611..360A, 2016MNRAS.457.3593F, 2016arXiv160902153H}. The dynamical importance of gas--grain chemistry in cooler regions of the disc is currently yet to be determined, for example presently unidentified chemically induced dynamical instabilities could potentially arise (see the burning questions, section \ref{burningQs}).

Aside from the coupling of chemistry with new physics such as dynamics, it is very important to stress that our base understanding of astrochemistry is constantly and rapidly evolving, with new species, reactions and regimes being identified that can only be studied in a \textit{dedicated} manner \citep[for example][use 10,000 models to study the sensitivity of single point chemical models to binding energies]{PenteadoInPrep}. It is important that such focused study continues.

Considering again the dust, there is no obvious consensus at present as to the best way to perform self-consistent dusty radiation hydrodynamics calculations of protoplanetary disc evolution. Schemes such as the short characteristics Variable Eddington Tensor (VET) method implemented in the {\sc Athena} code by \citet{2012ApJS..199....9D}, or the hybrid approach by \citet{2013A&A...555A...7K} show promise for bridging the gap between FLD and ray-tracing, but still require accurate modelling of the small grain dust population to determine the opacities before they can be applied in the context of protoplanetary discs (see Section~\ref{sec:dustdyn}).

With respect to magnetic fields, there are now also some approaches capable of modelling both radiation and magneto-hydrodynamics \citep[e.g.][]{2013A&A...560A..43F, 2015ApJ...801..117T}

Based on the above, we are already making excellent progress in cross-disciplinary modelling of discs, but most of this progress is very recent. There are still a number of highly coupled processes that cannot yet be modelled. As we will now discuss, there is a long, but fruitful journey ahead of multiphysics disc modellers.

\section{BRIDGING THE GAPS - CHALLENGES}
\label{sec:challenges}
The interconnectedness of different processes in discs means that to be able to answer many of the outstanding theoretical and observational questions regarding protoplanetary discs we will require a combination of three-dimensional, global, multi-phase simulations with radiation hydrodynamics, dust dynamics and size evolution, and chemistry computed self-consistently  (see Figure \ref{fig:connections}). 

\subsection{Burning questions}
\label{burningQs}
Some examples of `burning' science questions raised either during our discussion sessions, or by members of the community commenting on this manuscript, which might motivate improved multiphysics modelling of discs, included:

\begin{itemize}

\item What are the main drivers of global disc evolution? In particular, what is the main driver of the mass accretion rate in protoplanetary discs?
\item Alongside magnetic fields, what other processes govern or control the launching of jets and outflows?
\item{What is the effect of environment on protoplanetary disc evolution? For example, discs close to O stars are clearly heavily disrupted by high energy photons (we observe such systems as proplyds), but what is the role of comparatively modest radiation fields?}
\item{Do chemical--dynamical instabilities exist, i.e. is there a chemical reaction that feeds back into the dynamics (e.g. thermally) but responds to the dynamical change with a faster reaction rate?}
\item What happens to small grains at the surface of the disc or in outflows/winds?
\item What happens at high dust to gas ratios? How important are streaming instabilities, or other instabilities? How important are self--induced dust traps? What happens to dust in shocks?
\item How do magnetic fields in the disc affect the behaviour of charged dust grains, and how do the dynamics and ionisation chemistry of the grain population in turn affect the magnetic field evolution?
\item What are the conditions under which pebble accretion (e.g. \citealt{2010A&A...520A..43O,2012A&A...544A..32L,2012A&A...546A..18M}) might operate, and how will this impact the diversity of planetary systems formed in protoplanetary discs \citep[e.g.][]{2015A&A...582A.112B,2016ApJ...825...63C,2016A&A...591A..72I}? 
\item What is the nature of fragmentation in self-gravitating discs? Is there a well-defined parameter space where fragmentation occurs \citep[cf][]{Meru2011,Michael2012,Rice2012,Rice2014}, or can it occur stochastically through rare high-amplitude density perturbations over long enough timescales \citep{Paardekooper2012,Young2016}?
\item What is the origin of rings, gaps, horseshoes and cavities observed in mm-continuum emission? How common are these features?
\item How can the masses and properties of embedded protoplanets be constrained from observations? 
\item How do planets affect observations of chemical tracers?
\item{How do planets and circumplanetary discs affect the evolution of the protoplanetary disc (e.g. through thermal feedback or increased radiative heating in gaps). Conversely, how does the disc affect an embedded planet (e.g. the planetary atmosphere).}
\item Will dust discs fragment?
\item What determines the scale height of the dust layer? How is this set by different processes, for example, coagulation \citep[e.g.][]{2016ApJ...822..111K}
\item Under which conditions do warps develop in discs? Can radiation pressure drive warping?
\item{What are the possible initial conditions of class I/II/III discs and how do they influence the subsequent evolution? In particular how does the early evolution of discs affect the chemistry and grain distribution \citep[e.g.][]{2014A&A...567A..32M}? What is inherited from the star formation process?}
\item{The vertical component of the magnetic field controls the mass flux of winds and the saturation level of MRI-driven turbulence. How does the competition between accretion (drawing the vertical field in towards smaller radii) and diffusivity (pushing it outwards towards larger radii) cause this component of the field to vary with time? In particular what is the magnitude of the diffusivity term, which is set by microphysics  \citep[e.g.][]{1994MNRAS.267..235L, 2008ApJ...677.1221R, 2014ApJ...797..132T}?}
\item{How turbulent are protoplanetary discs \citep[e.g.][]{2015ApJ...813...99F, 2016A&A...592A..49T}?}
\item{What is the process by which a protoplanetary disc becomes a debris disc? Transition discs; those with inner holes, are typically attributed to the action of photoevaporation by the host star \citep[see e.g.][]{2016PASA...33....5O}, or planets \citep[e.g.][]{2011ApJ...729...47Z}. But which, if either, of these is the dominant process \citep[examples of models including both are][]{2009ApJ...704..989A, 2015MNRAS.454.2173R}? Are there other processes that contribute significantly to disc dispersal, such as magneto--thermal winds \citep{2016ApJ...818..152B}? What are the initial conditions of debris disc models \citep[e.g.][]{2005ApJ...627..286T, 2015PASA...32...39T}? }
\end{itemize}

Some of these questions might only be addressed by combining all of the physical ingredients of protoplanetary disc modelling.  However, several will only require consideration of a smaller fraction.  These smaller steps will be extremely valuable in bridging the gaps between fields, and will undoubtedly inform the production of a fully comprehensive modelling approach.  We manifest these steps as a series of challenges, outlined below.

\subsection{Grand challenges for gas modelling}

\subsubsection*{C1: Model the pressure and temperature effects of photochemistry in multidimensional, fully hydrodynamic models}

This challenges us to account for the (non--hydrostatic) dynamical impact of gas whose composition and temperature is set by photodissociation region processes. Specifically, the temperature should be accurately computed to within $\sim15$ per cent of a standard PDR network (which is the level of accuracy typically attained by reduced networks, see section \ref{sec:reduced}).


\subsubsection*{C2: Model the pressure and temperature effects of gas-grain chemistry in multidimensional, fully hydrodynamic models}

Similar to challenge C1, this challenges us to account for the (non--hydrostatic) dynamical impact of chemical processes in optically thick regions of discs. There is a caveat to this challenge in that the dynamical importance of gas-grain chemistry is currently unknown.  This therefore also (first) challenges us to determine what features of gas--grain chemistry might actually be dynamically important -- such as chemically induced dynamical instabilities (see also the burning questions; section \ref{burningQs}). 



\subsubsection*{C3: Incorporate the radiation field self-consistently while computing a multidimensional hydrodynamic model which satisfies challenges C1/C2}

Challenges C1 and C2 are likely to be met by making simplifying assumptions about the incident radiation and cosmic ray background.  The next step is then to properly account for the radiation field: set by the central protostar, the disc material and any surrounding environment (e.g. the envelope or neighbouring stars/clouds/associations).  This challenge will play a crucial role in understanding environmental influences on disc lifetimes.


\subsubsection*{C4: Model magnetic fields that can couple self-consistently to a realistic population of participating species}

Models constructed to meet challenges C1-C3 that directly compute the composition of matter will deliver self-consistent populations of electrons, ions and neutral species.  The formation and evolution of magnetically active and dead zones, and the activation of MRI, is fundamental to the disc's ability to accrete onto the star, as well as the launching of jets and outflows.  We must therefore be able to couple the magnetic field evolution to the above gas-grain chemistry (see also challenge C9).  Typically, MHD simulations that model the principal non-ideal processes (the Hall effect, Ohmic dissipation and ambipolar diffusion) use simplified models for ion/grain mass and charge, often assuming single values for these properties.  In practice, ion masses and charges will vary tremendously depending on the gas composition and the ambient radiation field. 

In this challenge, non-ideal MHD models must be made flexible enough to accept arbitrary populations of a wide variety of ions (and grains, see C9) as input for computing subsequent magnetic field structure \citep[c.f. the recent use of a reduced network by][]{2015ApJ...801..117T}.

\subsubsection*{C5: Assemblage of gas modelling challenges }
This essentially challenges us to model all components of the gas phase, i.e. to couple both C1 and C2, while incorporating C3 and C4. This challenge has two tiers. The lower tier involves accounting for all of the dynamical effects, without necessarily directly modelling the composition. Conversely the higher tier does involve direct computation of the dynamically (and observationally) relevant chemical species.

\subsection{Grand challenges for dust-gas modelling}

\emph{Simultaneously compute the dynamics and size evolution of the entire grain population, coupled to self-consistent modelling of the gas and radiation field in the disc in global, 3D simulations.} This can be broken into a series of smaller challenges, as follows:

\subsubsection*{C6: Model the dynamics of the entire grain population in a global disc simulation}
Develop the means to accurately and efficiently model the dynamics of solids spanning an entire grain size distribution in global, three dimensional, disc simulations, including the effect of embedded companions and with feedback from the dust grains to the gas.

\subsubsection*{C7: Model the growth and fragmentation of solids}
Develop an accurate prescription for growth and fragmentation of grains and incorporate it into 3D dynamical models of dust and gas evolution in global disc, with feedback from the dust grains to the gas.

\subsubsection*{C8: Radiative equilibrium and radiation pressure}
Compute the radiative equilibrium temperature, as well as the radiation pressure force, in global 3D dynamical protoplanetary disc simulations, using multi-frequency radiative transfer.

\subsubsection*{C9: Coupling to MHD}
Allow the dust grain population, along with the radiation field, to determine the ionisation chemistry in the disc and use this to self-consistently model the development of jets, outflows and MRI turbulence in both local and global disc models

\subsubsection*{C10: Assemblage of dust modelling challenges}
Similar to C5, this challenges us to combine \textit{C6}--\textit{C9}. That is, to have a method of computing the motions of a whole grain distribution, including the evolution of grain sizes and the effects of radiation and magnetic fields.

\subsection*{C11: The grandest challenge (in this paper)}
\emph{Develop a single model capable of reproducing multi-tracer, resolved, observations of a given protoplanetary disc.} That is, perform a global disc simulation that solves for the gas and dust dynamics, as well as the dust and chemical evolution of the disc, that then predicts (to within a reasonable degree of accuracy) all observed properties of a given disc at a resolution comparable to that of current observational instrumentation. The model should retrieve the continuum morphology and intensity  for wavelengths probing a range of grain sizes, whilst also reproducing molecular line observations of different tracers (for example C$^{18}$O, HCO$^+$, $^{12}$CO, which probe different components
of the disc and can be sensitive to different chemical effects).

Doing so will require simultaneous completion of many of the above challenges. It is therefore a long term goal, but one which should be achievable given progress made on the other challenges stated above.

\begin{table*}
  \caption{A qualitative summary of the effect of different components of disc modelling on the intrinsic physical properties of protoplanetary discs -- ``\tickmark'' implies that an ingredient is identified as important, ``?'' implies that the importance is uncertain, ``\crossmark'' implies that an ingredient is likely unimportant.  It is our hope that such a summary would eventually become more quantitative, with the relative importance of different processes more formally assessed.}
  \label{egTable}
  \begin{tabular}{cr|cccccccc}

  & & \multicolumn{5}{c}{PROCESSES} \\
  
   &   & Accretion & Planet  & Winds & Disc dispersal/ & Jets/ &   Observations & \ldots \\
   & &  &formation & &lifetimes  & Outflows & \\
 \hline
 \parbox[t]{2mm}{\multirow{10}{*}{\rotatebox[origin=c]{90}{INGREDIENTS}}} & Hydrodynamics & \tickmark & \tickmark & \tickmark & \tickmark & \tickmark& \tickmark &  \ldots \\
                     & Self-gravity & \tickmark & \tickmark & \crossmark & ?& ? & \tickmark &  \ldots \\
                     & Dust dynamics &  ? & \tickmark & ? &? & ? &\tickmark &  \ldots \\ 
                     & Magnetic fields &  \tickmark & ? & \tickmark & \tickmark & \tickmark &\tickmark &  \ldots \\
                     & Radiation transport &  ? & \tickmark & \tickmark & \tickmark & \tickmark &\tickmark &  \ldots \\
                     & (Proto)-Stellar Evolution &  \tickmark & ? & \tickmark & \tickmark & \tickmark &\tickmark &  \ldots \\
                     & Photoionisation &  ? & \crossmark & \tickmark &\tickmark & \tickmark & \tickmark &  \ldots \\
                     & PDR chemistry &  ? & \crossmark & \tickmark &\tickmark & ? & \tickmark &  \ldots \\
                     & Gas-grain chemistry  &  \crossmark & \tickmark & \crossmark & \crossmark & ? &\tickmark &  \ldots \\
                     & \vdots       &     \vdots          & \vdots   &    \vdots        &   \vdots         & \vdots &      \vdots       & \\     
\end{tabular}
\end{table*}

\section{DISCUSSION - STRATEGIC STEPS TOWARDS THE FUTURE}
\label{Strategy}

The grand challenges discussed in the previous section are in a sense a strategic pathway towards long term future development. In practice models of discs are currently much more focused, but could still be improved by the integration of previously uncoupled physics. In this section we discuss broad strategy for the immediate future of more general disc modelling. More specific technical developments are discussed in the next section.

\subsection{Which problems are the most pressing to solve and what physics is required to solve them?}

It is inefficient to develop new software, or exhaust substantial CPU hours on an intensive state of the art multiphysics calculation, if the results have no value. A key strategic step, therefore, is to identify the combination of physics required to answer well motivated, well formulated, key problems.

\smallskip

Table \ref{egTable} provides an example of a strategic overview. Such an overview can guide/motivate the development of numerical methods to include all of the physics essential to solve a given problem. It would also motivate us to understand whether the uncertain features really do play an important role. 

\smallskip

In addition to identifying the processes that might contribute to a problem (such as in Table \ref{egTable}), one could possibly then order the contributing physical processes in a hierarchy of importance to determine which are the most important features to include in a model \citep[similar to the way that the dynamical importance of microphysics on H\,\textsc{ii} region expansion was categorised by ][]{2015MNRAS.453.2277H}. For example, consider the generation of synthetic molecular line observations. At the most basic level radiative transfer is required, as it is photons that are observed by astronomers, as well as some estimate of the density, temperature, molecular abundance and molecular level populations. This can initially be done assuming some simple static disc structure, assuming an abundance of molecules and level populations determined analytically by the Boltzmann distribution. This could then be improved with proper non local thermodynamic equilibrium (NLTE) statistical equilibrium calculations, which could be improved upon by using chemical networks/direct abundance calculations, which can be improved upon by further solving the dynamics/thermal balance, decoupled dust transport and so on. 
In order to do this one would first need to define some measure of importance. For example if interested in accretion a hierarchy of importance would place processes resulting in the largest contribution to the accretion rate at the top.

Deciding which problems are most pressing to address should also be informed by recent and upcoming observations. For example, which questions might be addressed by models in tandem with data from the Square Kilometer Array \citep[SKA, which among other things will probe grain growth and disc chemistry][]{2015aska.confE.117T}, James Webb Space Telescope (JWST) or the European--Extremely Large Telescope \citep[E--ELT, e.g.][]{2009AIPC.1158..333H}? 

Another key strategic point is to stress that on the path towards multiphysics modelling of {significantly interdependent physics} it is essential that all individual fields continue to develop as they are presently. Integration should be \textit{complementary} to our current approaches. There are (at least) two key reasons for this. One reason is that an integrated approach is likely to be substantially more computationally expensive, which limits the parameter space of a given problem that can be studied. This also strongly limits the ability to quickly interpret observations (e.g. with parametric models). The other reason is that each field is continuing to evolve, with the development of new algorithms and the identification of new important mechanisms. This focussed field-by-field progression will likely answer a number of the burning questions and the techniques developed will ultimately feed back into multiphysics models of the future.

\section{TECHNICAL STEPS TOWARDS THE FUTURE}
\label{Technical}

We now discuss possible near-term developments of our numerical methods towards resolution of the grand challenges, focussing on the coupling of physical ingredients with a particular emphasis on chemistry. 

\subsection{Simplified chemistry for dynamics}
We currently identify three possible approaches to including chemistry in dynamical simulations: direct calculation of a full network and heating/cooling rates, direct calculation of a reduced network, or implementation of pre-computed look-up tables.  We discuss these further below   

\subsubsection{Reduced chemical networks}
\label{sec:reduced}
Reduced chemical networks prioritise only the species and reactions of most importance to a given aim. For example, if prioritising dynamics, then an ideal reduced network would be one that yields a temperature/pressure to within an acceptable degree of accuracy (say 10--15 per cent). The established method of generating a reduced network is to start with a comprehensive one and systematically remove components, checking that it does not have a substantial impact on the resulting quantity of interest. There are already codes available capable of computing chemistry based on very large networks, such as \textsc{prodimo} \citep{2009A&A...501..383W}, \textsc{dali} \citep{2013A&A...559A..46B}, \textsc{ucl-chem} \citep{2004MNRAS.354.1141V, 2011ApJ...740L...3V}, \textsc{ucl-pdr} \citep{2005MNRAS.357..961B, 2006MNRAS.371.1865B} and the models of \cite{2012ApJ...747..114W}.  Any of these networks could be analysed to determine which processes are essential for dynamics, and then reduced accordingly.  Additionally, it is also possible to optimise calculations of large networks \citep[e.g.][]{grassi_2013}.  It is likely that a combined approach of reduction and optimisation will yield the most efficient results.  

\smallskip

PDR chemistry is important in surface layers and the disc outer edge if the disc is externally irradiated. Reduced PDR networks already exist \citep[e.g.][]{2007A&A...467..187R}. Such a network is already used in dynamical models by \textsc{torus-3dpdr} (see section \ref{SOAhybrid}). However existing reduced PDR networks are predominantly motivated by studies of star forming regions/the interstellar medium. New reduced networks tailored for discs would be extremely valuable for future dynamical models including PDR chemistry.  

\smallskip

In the same vein as reduced chemical networks, there are also some recent promising developments concerning the relatively computationally cheap determination of the ionisation state in dense, dusty, optically thick regions of discs (in particular where dust-phase recombination dominates over the gas-phase) which is particularly important for MHD and evolution of the dust population (e.g. regarding coagulation). \cite{2016arXiv160703701I} present an analytic model that yields the ionisation state of such dusty media, which could be incorporated into non-ideal MHD codes -- offering an imminently achievable significant advance. 

\subsubsection{Lookup tables and functional parameterisations}
\label{sec:lookup}

An alternative to direct computation of the chemistry/temperature using a ``full'' or reduced network is to tabulate temperatures or heating/cooling rates as a function of local properties in a disc. For example, \cite{2010MNRAS.401.1415O} prescribe the temperature of gas optically thin to X--rays as a function of local ionisation parameter (i.e. the density, distance from the source and stellar X--ray luminosity) where the function \citep[itself only published in full in][]{2016MNRAS.457.1905H} was computed by the dedicated photoionisation code \textsc{mocassin} \citep{2003MNRAS.340.1136E, 2008ApJS..175..534E}. A similar approach to obtaining PDR or gas-grain chemistry temperatures, where lookup tables are computed prior to run-time, could vastly reduce the potential computational expense of dynamical models. 

Unfortunately, chemistry (both gas-grain and PDR) is not generally so easily parametrised as a simple function of the local properties.  In order to include all relevant effects of heating and cooling, such a look-up table could easily grow very large.  Below we briefly list several example quantities that would need to be included, along with a typical dimensionality for each in parenthesis (I.~Kamp, private communication):

\begin{itemize}
    \item The temperature of dust grains (1).
    \item The dust grain size(s), including second moment of the size distribution for grain surface chemistry and collisional gas-grain coupling (2). 
    \item The dust grain density (1).
    \item The gas density (1).
    \item Column densities towards the central star of key species (H, C, CO) for evaluating the amount of shielding (3).
    \item The cosmic ray ionisation rate (though this can perhaps be approximated as constant throughout the disc) (1).
    \item The strength of the radiation field in several bands, including X-Ray, UV and optical (10).
    \item The optical depth of the dust in direction of closest escape (1).
    \item The fractional abundance of polycyclic aromatic hydrocarbons (PAHs) and further details if also using PAHs as an opacity source (3). Again, these parameters may be constants throughout the disc.
    \item Column densities of all species to be considered, both toward the radiation source, and the direction of closest escape ($\gtrsim$10). 
\end{itemize}

Given that the above list is by no means exhaustive, it is easy to see that such a look-up table may reach a dimensionality of 30--40.  One of the key factors accounting for this issue is that the local chemistry depends upon the 3D non--local density distribution, because this sets photon escape probabilities, i.e. the chemistry at some point in space cares about the gas distribution in all directions from that point. It is therefore not solely dependent upon local properties, even if the local radiation field from external sources has been computed.

However, many of these quantities are likely not entirely independent, and relations between them could be identified in a statistically robust manner using grids of simulations.  This may allow a reduction in the number of dimensions required. Of further note is that a ``simplified'' thermodynamic prescription based on chemical modelling was developed by \cite{1996A&A...311..927W,1996A&A...313..217W, 2003A&A...404..267S} for application to pulsating stars, which might offer some guidance on how to streamline some of the aforementioned dependencies.

\subsection{Direct hybridisation}
\label{SOAhybrid}


Historically the approach to including more physics in dynamical models is to use a hydrodynamical code as the foundation and incorporate simplified physics modules subsequently. For example \cite{2010MNRAS.404....2G} and \cite{2016arXiv160508032D} patch reduced chemical networks into \textsc{zeus-mp} and \textsc{ramses} respectively. \cite{2013A&A...560A..43F} also present an extension of the \textsc{pluto} code that includes both magnetic fields and an FLD radiation transport scheme. There is another approach, which is to start with a state of the art chemistry/radiative transfer code and subsequently incorporate somewhat more simple hydrodynamics. An example of this latter approach is the \textsc{torus} radiation transport and hydrodynamics code. This code began its life solely as a Monte Carlo radiative transfer code \citep{2000MNRAS.315..722H} but now includes hydrodynamics, so can perform radiation hydrodynamic simulations with all the features of a dedicated radiation transport code \citep[e.g. detailed photoionisation, dust radiative equilibrium and radiation pressure in arbitrarily complex system geometries, etc.;][]{2012MNRAS.420..562H, 2015MNRAS.448.3156H, 2015MNRAS.453.2277H}. Furthermore, \textsc{torus-3dpdr} is an extension of \textsc{torus} that also includes PDR chemistry \textit{with 3D extinction and escape probabilities} \citep{2015MNRAS.454.2828B}.  The UV radiation field everywhere is computed using the Monte Carlo radiation transport and the escape probabilities are estimated in 3D using an algorithm based on \textsc{healpix} \citep{2005ApJ...622..759G}. \textsc{torus-3dpdr} is capable of directly modelling the role of far ultraviolet (FUV) photons dynamically in non-hydrostatic scenarios, such as the external irradiation of discs by FUV radiation that has only been possible semi-analytically in the past \citep[][]{2004ApJ...611..360A, 2016MNRAS.457.3593F, 2016arXiv160902153H}. It could also be used to test the validitiy of escape probability methods that assume a single dominant trajectory (the 1+1D methods).

\smallskip

One argument in favour of adding hydrodynamics to a radiative transfer/chemistry code is development time, since a simple but effective hydrodynamics algorithm is usually much more straightforward to develop than a radiative transfer/chemistry algorithm (though of course care must be taken to ensure that the hydrodynamics algorithm is appropriate for any given application). The obvious argument against this coupling of state of the art physics models with hydrodynamics is that they are not necessarily well streamlined and can be very computationally expensive \citep[though this is not necessarily a problem if the code is optimised and/or highly scalable, as is the case for \textsc{torus},][]{2015MNRAS.448.3156H}. 



\smallskip

Constructing a dedicated photochemical-dynamical code from scratch is another possible option, but potentially requires a lot of development time \citep[e.g. the recent PDR-dynamical code of][]{2015ApJ...808...46M}.

\smallskip

Another promising avenue is the development of diverse, flexible self-consistent physics libraries that can be ported into other numerical (and therefore potentially hydrodynamical) codes. The \textsc{krome} code is an excellent example of this approach, which quickly solves arbitrary chemical networks and can also calculate heating and cooling terms \citep{2014MNRAS.439.2386G}. Spectral codes, which solve partial differential equations flexibly and efficiently, could also offer a powerful means of combining other physical ingredients in a relatively straightforward manner. Spectral codes appear not to have featured in multiphysics disc modelling to date, but options for doing so include the \textsc{dedalus} \citep{2016ascl.soft03015B} and \textsc{snoopy} \citep{2015ascl.soft05022L} codes.

\subsection{Temporal and spatial resolution}
\label{sec:ResTime}
A very specific problem is that (in particular for non-equilibrium chemical-dynamics) we have to determine what the spatial and temporal scales are that we have to resolve in a given scenario. As an example, chemical timescales in the disc upper layers (that is, in the PDR regions) are rather short, whereas timescales deeper in the disc are usually much longer (for example, the case of CO being converted into CH$_{4}$ on timescales even longer than protoplanetary discs lifetimes). The time steps required to model the upper layers may therefore eventually be limited by the chemical timescales (in non-equilibrium scenarios) rather than the dynamical timescales, which might drastically increase computational expense. In such a regime where the chemical timescale is very small (much smaller than the dynamical timescale) we may be able to alleviate the problem somewhat with chemical sub-stepping - running multiple chemical updates per hydrodynamic update. Conversely if the chemical/thermal timescales (reaction/heating/cooling rates) are very long, many dynamical steps can be taken between the more expensive chemical updates, improving the computation time substantially. 

Alternatively, if the system is expected to reach a steady state, and all that is desired is an accurate model of this steady state (rather than the pathway to reaching the steady state) it may be possible to run chemical calculations very infrequently even if the chemical timescale is very short.

\smallskip

In addition to the above timescale arguments, resolution also needs to be considered. For example some chemical features may only arise if the spatial resolution (e.g. around shocks) is sufficiently high - capturing such processes will of course increase computational expense.

\subsection{Scaling}
A key technical consideration is the scaling of the various physical ingredients in terms of both elements (cells, rays, chemical species, reactions etc.) and computational resources (number of cores), since a calculation is going to be limited by its least tractable component. Different numerical approaches to computing a given ingredient can scale very differently. For example, in the case of radiative transfer, Monte Carlo radiation transport and \textsc{treecol} \citep[see][for details of the latter]{2012MNRAS.420..745C} scale much more efficiently than long characteristic ray tracing. There are therefore multiple scaling options per ingredient. 

\smallskip

For applications comprising two or more ingredients that scale very differently, there will likely be idle cores/inefficient CPU usage in the components of the code that do not scale so well. Furthermore some techniques have specific constraints on the number/configuration of cores which may vary for different calculation ingredients. For example if the hydrodynamic component of a calculation were confined to $i$ distributed memory MPI threads (plus an arbitrary number of shared memory openMP threads), but the radiative transfer to $j > i$ MPI threads, there will be unused MPI threads during each hydrodynamics step. This is a situation where dynamically optimising between shared and distributed memory processes is worthwhile, setting the otherwise idle MPI threads to contribute to openMP or other useful tasks. 

\subsection{Hardware developments}
It is also important to assess new and projected hardware developments. We are approaching a time in which access to large numbers of processors increasingly outweighs the developments in performance of the processors themselves. Efficiently scalable numerical methods, such as Monte Carlo radiation transport and discontinuous Galerkin hydrodynamics solvers, will therefore be extremely advantageous in the near future. 

Another significant realisation (only recently for astronomers) is that graphics processings units (GPUs) can offer significant speedup per core. A relatively small (but growing) fraction of astrophysical codes have a GPU implementation, and those that do are often those used for cosmological applications \citep[e.g.][]{2010ApJS..186..457S, 2014ApJS..211...19B}. However, a GPU implementation of the \textsc{fargo} disc-modelling code was developed by \cite{2016ApJS..223...11B}, where they quote a typical speedup per core of a factor 40. It is beyond the scope of this paper to discuss GPU programming in detail, but we note that GPUs are fundamentally different architectures to CPUs and are therefore programmed in a somewhat different manner (taking time to learn), typically using either the \textsc{cuda} \citep{Nickolls:2008:SPP:1365490.1365500} or \textsc{opencl} \citep{stone2010opencl} standards. The high speeds of GPUs make them a powerful tool for the future of astronomy, where applicable, and they are likely to feature much more frequently in astronomy in the coming years, especially with the advent of directive-based GPU acceleration using the OpenACC standard\footnote{\url{http://www.openacc.org}}.

A final example, mentioned here only in passing, is the introduction of new types of processor such as the Intel Xeon Phi \citep[e.g.][]{jeffers2013intel} - which combines some of the performance advantages of GPUs with an easier programming framework. 

In general the writing of new codes, or adapting old ones, to take advantage of hardware developments will be important. Given that more specialised hardware might continue to appear over time, it would also be advantageous if codes could be developed in such a fashion that they are easily ported, but it is unclear (to us at least) exactly how this might work in practice. This is an area where increased collaboration between astrophysicists and computer scientists will be advantageous. Interaction with computer scientists could also lead to other benefits such as improved efficiency of our codes and the promotion of better coding practice.   

\section{COLLABORATIVE STEPS TOWARDS THE FUTURE}
\label{Collaboration}

As already mentioned, the components that we want to couple in the future of disc modelling are themselves already established and complex fields.  It is therefore clear that these challenges are a whole-community effort, and substantial progress will only be made via collaboration.  To this end, we have identified several key collaborative steps that we discuss below.

\subsection{Workshops}
Workshops are likely to be essential for stimulating cross-disciplinary collaboration.  While a typical conference setting will be important for each sub-discipline to discuss their work generally, events with ample time for break-out sessions and collaborative spaces are likely to be very productive.  Such events allow large-scale discussion, but also allow for specific problems to be tackled one-on-one or in small groups in an `unconference' setting (for example, the dotAstronomy\footnote{\url{http://dotastronomy.com/}} or Astropy\footnote{\url{http://www.astropy.org/}} conference series).  The identification of key ingredients to be swapped between respective fields will be important to establish, e.g.\ heating and cooling rates are likely to be of interest to those running dynamic models, while detailed abundance results may not be required.

\subsection{Benchmarking}
In addition to workshops, it is important for each field to develop an agreed set of benchmark problems, with the aim of transparency and reproducibility. Code comparison projects are key, but can require a lot of work for a small number of publications \citep[albeit high impact, e.g.][]{2006MNRAS.370..529D, 2007A&A...467..187R, 2009A&A...498..967P, 2009MNRAS.400.1283I}.  

A good example of a successful comparison project is the recent StarBench code comparison workshops\footnote{\url{https://www.astro.uni-bonn.de/sb-ii/}} \citep{2015MNRAS.453.1324B}.  These workshops aimed at testing radiation hydrodynamics codes used to study problems in star formation, with an emphasis on doing so in a positive and friendly environment. The workshops involved attendees running tests before arrival, which spanned a range of complexity. In the first meeting at the University of Exeter in 2013, every code passed the purely hydrodynamic shock tests without issue. However the instant that radiative transfer/photoionisation was introduced into the dynamical problem we generally had poor agreement, even for the simplest case of tracking the time evolution of the extent of an ionised region about a star in a uniform density medium composed solely of hydrogen. The origin of the inconsistency between codes was that they were all running slightly different models (e.g. inconsistent recombination rates) and, after extremely careful rewriting of the specifications of this simple test, were subsequently able to get truly excellent agreement between the codes. \textit{This process highlighted to the community all of the things that should be explicitly stated in a paper in order to make it truly reproducible}. Last but not least, in the case of an expanding H\,\textsc{ii} region we actually discovered that although the codes all agreed perfectly, they did not agree with the classic analytic solution that everyone would compare with in their numerical methods paper and suggest that they get ``good enough'' agreement with --- validating their approach. Following re-investigation, as a result of code comparison, a direct improvement in our understanding of this fundamental analytic problem has been established \citep{2015MNRAS.453.1324B}.   In summary, code comparison 
\begin{itemize}
    \item Verifies that codes are working as desired
    \item Informs the community what needs to be specified in papers to make them reproducible --- a key factor, especially since there are likely to be many more ingredients in disc models of the future than there were in the relatively straightforward StarBench tests. 
    \item Improves our understanding of each other's numerical methods, including relative strengths and weaknesses. This can be done in a friendly way.
    \item Highlights the importance of careful numerics (e.g. understanding resolution dependency and which techniques are appropriate for a given scenario).
    \item Results in high impact publications.
    \item Leads to an improvement in our understanding of the underlying more fundamental (even analytic) problems. 
\end{itemize}
Key to a successful comparison is \emph{active feedback between participants} and \emph{iteration towards understanding} the origin of any differences encoutered. This can often be achieved just as easily with a comparison involving just two or three codes performed by a relatively small team (e.g. \citealt{1997MNRAS.288.1060B,2008A&A...482..371C,2010MNRAS.406.1659P,2013MNRAS.432..711H}). Such an approach avoids much of the friction associated with large-scale comparison projects while achieving the same objectives.

\subsection{Open source software}
A more applied collaborative practice is to develop software in an open source format (e.g. using GitHub\footnote{\url{https://github.com/}}). This is potentially very useful for both transparency and distributed development (i.e. international contributors). Examples taking such an approach are the \textsc{krome} \citep{2014MNRAS.439.2386G}  and \textsc{lime} \citep{2010A&A...523A..25B} projects.

Although the open source mentality is desirable, it should not be imposed since there may be legitimate reasons to protect intellectual interests. For example, if an early-career researcher invests substantial time into code development, the current academic culture requires a period where they are able to see a return on their time investment, in terms of first author publications where they lead astrophysical research (in the current culture, this \textit{is} more important than a number of co-authored publications). There is no reason that their code cannot be shared collaboratively during such a phase of research. More widespread access can subsequently be yielded once the developers have seen sufficient return.

\section{SUMMARY}
Protoplanetary discs are a key focus of modern astronomy, being subject to extensive modelling including the dynamics of gas and dust, magnetic fields, radiation transport and chemistry. These facets of physics required to model discs are, however, not independent, so as we proceed into the future we must consider their coupling in multiphysics modelling of discs. In particular, we perceive that it will be important to self-consistently model decoupled gas and dust dynamics, with radiative transfer, dust growth/fragmentation and different chemical regimes (gas-grain, PDR). This paper aims to stimulate this development and consisted of the following components. 

Firstly, to establish a platform from which to discuss the coupling of different disciplines, we provide an overview of each in isolation, as well as the progress made towards multiphysics modelling to date. Using this, we have identified a series of challenges for the future of protoplanetary disc modelling, which are supposed to act as milestones towards the ultimate goal of a self-consistent gas, dust, radiation transport and chemistry model mentioned above. Our first category of challenges regards gas modelling, with a particular focus on composition (e.g. gas-grain and photochemistry) coupled with dynamics. Our second category of challenges regards dust, including modelling of an entire grain size distribution as well as growth and fragmentation of grains and any additional physics (such as radiation) that alters the dust dynamics. We also discuss pathways towards addressing these challenges, which are grouped by whether they are strategic (e.g.\ identifying what needs to be done), technical (e.g.\ working out how to do it) and collaborative (working together to do it). 

We finish by noting that, as a further motivational strategy, appropriate agents mights offer prize(s) for completing more rigorously defined versions of one or more of the challenges presented here.

\section*{ACKNOWLEDGMENTS}
We thank the anonymous referee for their comments, and positive review of the manuscript. We would like to acknowledge the Protoplanetary Discussions conference, including the members of the local and scientific organising committees.  We also thank the attendees of the conference (Table \ref{tab:participants}) and those who took part in the discussion sessions chaired by John Ilee and Daniel Price, all of whom helped inform many of the statements made here. We also particularly thank Phil Armitage for comments on the manuscript. 

Through most of this work TJH was funded by the STFC consolidated grant ST/K000985/1 and is now funded by an Imperial Junior research fellowship.  JDI gratefully  acknowledges  support  from  the  DISCSIM  project,  grant agreement  341137,  funded  by  the European  Research  Council  under ERC-2013-ADG.  DHF acknowledges support from the ECOGAL project, grant agreement 291227, funded by the European Research Council under ERC-2011-ADG. DJP gratefully acknowledges funding via Future Fellowship FT130100034 from the Australian Research Council.OP is supported by the Royal Society Dorothy Hodgkin
Fellowship

\bibliographystyle{pasa-mnras}
\bibliography{hydro_rt_chem}

\begin{table*}
 \centering
  \caption{Participants of Protoplanetary Discussions 2016}
  \label{tab:participants}
  \begin{tabular}{L{0.29\textwidth} C{0.29\textwidth} R{0.29\textwidth}}

  \multicolumn{3}{c}{\includegraphics[width=0.92\textwidth]{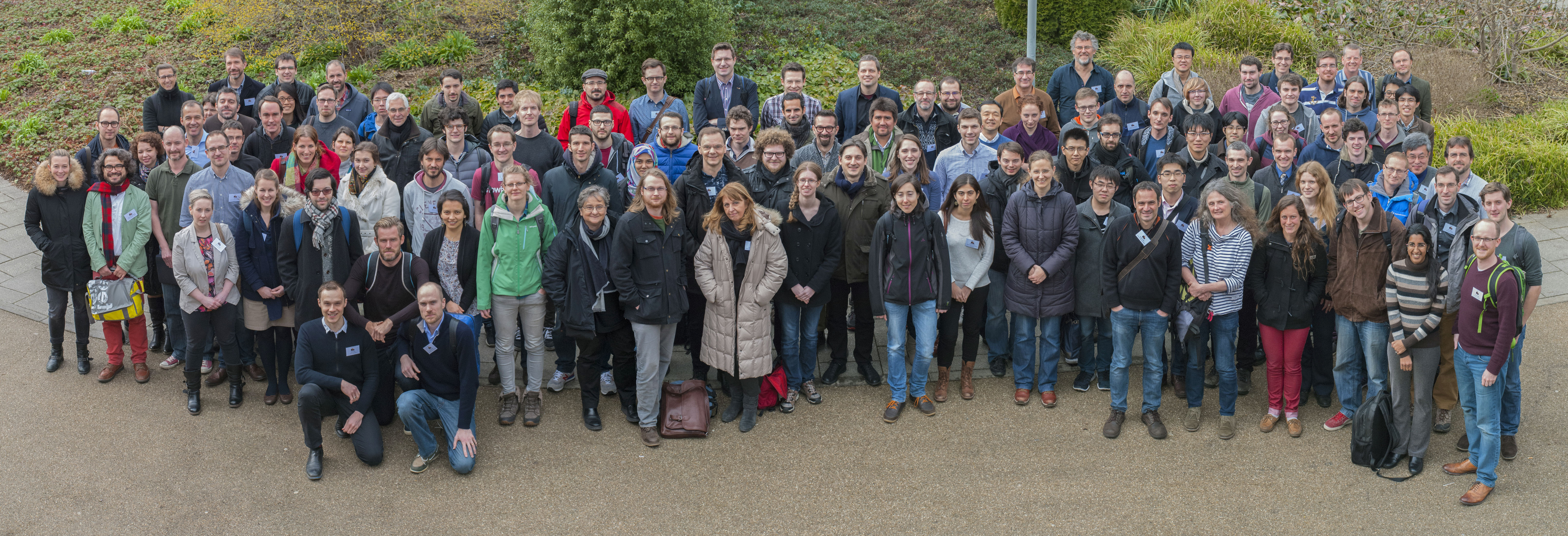}} \\

P{\'e}ter {\'A}brah{\'a}m	&	Jane Greaves	    &	Brunella Nisini	\\
Richard Alexander	        &	Aaron Greenwood 	&	Aake Nordlund	\\
Stefano Antonellini	        &	Oliver Gressel  	&	Shota Notsu	\\
Sareh Ataiee	            &	Viviana Guzm{\' a}n	&	Tomohiro Ono	\\
Francesca  Bacciotti	    &	Cassandra  Hall	    &	Ren{\'e} Oudmaijer	\\
Hans Baehr	                &	Tomoyuki Hanawa	    &	Olja Panic	\\
Andrea Banzatti	            &	Thomas Haworth	    &	Francesco Pignatale	\\
Matthew Bate	            &	Eyal Heifetz	    &	Paola Pinilla	\\
Myriam Benisty	            &	Michiel Hogerheijde	&	Christophe Pinte	\\
Olivier Bern{\'e}	        &	Edward Hone	        &	Adriana Pohl	\\
William Bethune	            &	Jane Huang	        &	Klaus Pontoppidan	\\
Dominika Boneberg	        &	Mark Hutchison	    &	Daniel Price	\\
Richard Booth	            &	John Ilee	        &	Christian Rab	\\
Jerome Bouvier	            &	Colin Johnstone	    &	Ken Rice	\\
Christian Brinch	        &	Attila Juhasz	    &	Donna Rodgers-Lee	\\
Claudio Caceres	            &	Mihkel Kama	        &	Giovanni Rosotti	\\
Hector Canovas	            &	Inga Kamp	        &	Vachail Salinas	\\
Andres Carmona Gonzalez	    &	Kazuhiro Kanagawa	&	Steph Sallum	\\
Mason Carney	            &	Lucia Klarmann	    &	Christian Schneider	\\
Paolo Cazzoletti	        &	Stefan Kraus	    &	Aurora Sicilia-Aguilar	\\
Jason Champion	            &	Guillaume Laibe	    &	KangLou Soon	\\
Cathie Clarke	            &	Geoffroy Lesur	    &	Tomas Stolker	\\
Ilse Cleeves	            &	Min-Kai Lin	        &	Hidekazu Tanaka	\\
Claire Davies	            &	Giuseppe Lodato	    &	Ryo Tazaki	\\
Odysseas Dionatos	        &	Ryan Loomis	        &	Richard Teague	\\
Thomas Douglas	            &	Pablo Loren-Aguilar	&	Jean Teyssandier	\\
Maria Drozdovskaya	        &	Carlo Felice Manara	&	Wing-Fai Thi	\\
Marc Evans	                &	Natascha Manger	    &	Amaury Thiabaud	\\
Stefano Facchini	        &	S{\'e}bastien Maret	&	Nienke van der Marel	\\
Kevin Flaherty	            &	Colin McNally	    &	Gerrit van der Plas	\\
Duncan Forgan	            &	Fran{\c c}ois M{\'e}nard	&	Riccardo Vanon	\\
Kevin France	            &	Ignacio Mendigut{\' i}a	&	Catherine Walsh	\\
Anthony Garcia	            &	Farzana Meru	    &	Matthew Willson	\\
Antonio Garufi	            &	Anna Miotello	    &	Peter Woitke	\\
Nikolaos Georgakarakos	    &	Takayuki Muto	    &	Ron Yellin-Bergovoy	\\
Jean-Fran{\c c}ois Gonzalez	&	Richard Nelson	    &	Zhaohuan Zhu	\\

\end{tabular}
\end{table*}

\end{document}